\documentclass[twocolumn]{aastex63}

\usepackage{hyperref}
\usepackage{graphicx}
\usepackage{natbib}
\usepackage{amsmath,amsthm,amssymb}

\def\WISE{\textit{WISE}}
\def\WD{\textit{WISE}/DEIMOS}

\protected\def\Lsed{\ifmmode \,\mathcal{L}_{\mathrm{SED}}\else $\mathcal{L}_{\mathrm{SED}}$\fi}

\protected\def\picometer{\ifmmode \,\operatorname{pm}\else $\operatorname{pm}$\fi}
\protected\def\nm{\ifmmode \,\operatorname{nm}\else $\operatorname{nm}$\fi}
\protected\def\micron{\ifmmode \,\operatorname{\mu m}\else $\operatorname{\mu m}$\fi}
\protected\def\mm{\ifmmode \,\operatorname{mm}\else $\operatorname{mm}$\fi}
\protected\def\meter{\ifmmode \,\operatorname{m}\else $\operatorname{m}$\fi}
\protected\def\km{\ifmmode \,\operatorname{km}\else $\operatorname{km}$\fi}
\protected\def\au{\ifmmode \,\operatorname{AU}\else $\operatorname{AU}$\fi}
\protected\def\pc{\ifmmode \,\operatorname{pc}\else $\operatorname{pc}$\fi}
\protected\def\kpc{\ifmmode \,\operatorname{kpc}\else $\operatorname{kpc}$\fi}
\protected\def\Mpc{\ifmmode \,\operatorname{Mpc}\else $\operatorname{Mpc}$\fi}
\protected\def\rsun{\ifmmode \,\operatorname{R_\odot}\else $\operatorname{R_\odot}$\fi}
\protected\def\Rsun{\ifmmode \,\operatorname{R_\odot}\else $\operatorname{R_\odot}$\fi}

\protected\def\second{\ifmmode \,\operatorname{sec}\else $\operatorname{sec}$\fi}
\protected\def\yr{\ifmmode \,\operatorname{yr}\else $\operatorname{yr}$\fi}
\protected\def\Gyr{\ifmmode \,\operatorname{Gyr}\else $\operatorname{Gyr}$\fi}

\protected\def\eV{\ifmmode \,\operatorname{eV}\else $\operatorname{eV}$\fi}
\protected\def\keV{\ifmmode \,\operatorname{keV}\else $\operatorname{keV}$\fi}
\protected\def\MeV{\ifmmode \,\operatorname{MeV}\else $\operatorname{MeV}$\fi}
\protected\def\GeV{\ifmmode \,\operatorname{GeV}\else $\operatorname{GeV}$\fi}
\protected\def\TeV{\ifmmode \,\operatorname{TeV}\else $\operatorname{TeV}$\fi}

\protected\def\Lsun{\ifmmode \,\operatorname{L_\odot}\else $\operatorname{L_\odot}$\fi}
\protected\def\lsun{\ifmmode \,\operatorname{L_\odot}\else $\operatorname{L_\odot}$\fi}
\protected\def\Watt{\ifmmode \,\operatorname{W}\else $\operatorname{W}$\fi}
\protected\def\nW{\ifmmode \,\operatorname{nW}\else $\operatorname{nW}$\fi}

\protected\def\kJy{\ifmmode \,\operatorname{kJy}\else $\operatorname{kJy}$\fi}
\protected\def\Jy{\ifmmode \,\operatorname{Jy}\else $\operatorname{Jy}$\fi}
\protected\def\mJy{\ifmmode \,\operatorname{mJy}\else $\operatorname{mJy}$\fi}
\protected\def\microJy{\ifmmode \,\operatorname{\mu Jy}\else $\operatorname{\mu Jy}$\fi}
\protected\def\nJy{\ifmmode \,\operatorname{nJy}\else $\operatorname{nJy}$\fi}

\protected\def\Mag{\ifmmode \,\operatorname{mag}\else $\operatorname{mag}$\fi}

\protected\def\deg{\ifmmode ^{\circ}\else $^{\circ}$\fi}
\protected\def\arcsec{\ifmmode ^{\prime\prime}\else $^{\prime\prime}$\fi}

\protected\def\arcsecT{\ifmmode \,\operatorname{arcsec}\else $\operatorname{arcsec}$\fi}
\protected\def\arcmin{\ifmmode ^{\prime}\else $^{\prime}$\fi}

\protected\def\arcminT{\ifmmode \,\operatorname{arcmin}\else $\operatorname{arcmin}$\fi}
\protected\def\sr{\ifmmode \,\operatorname{sr}\else $\operatorname{sr}$\fi}

\newcommand{\code}[1]{\texttt{#1}}

\protected\def\d{\ifmmode \operatorname{d}\else
    $\operatorname{d}$\fi}

\def\apjl{{ApJ}}

\def\apjs{{ApJS}}

\def\apjref#1;#2;#3;#4 {\par\pp#1, {#2}, #3, #4 \par}

\shorttitle{The $3.4\micron$ Extragalactic Background Light}
\shortauthors{Lake et al.}

\begin{document}

\title{The Contribution of Galaxies to the $3.4\micron$ Cosmic Infrared Background as Measured Using \WISE}
\author[0000-0002-4528-7637]{S.~E.~Lake}
\affiliation{National Astronomical Observatories, Chinese Academy of Sciences, Beijing, 100012, People's Republic of China}
\affiliation{CAS Key Laboratory of FAST, NAOC, Chinese Academy of Sciences, People's Republic of China}
\affiliation{Physics and Astronomy Department, University of California, Los Angeles, CA 90095-1547}

\author[0000-0001-5058-1593]{E.~L.~Wright}
\affiliation{Physics and Astronomy Department, University of California, Los Angeles, CA 90095-1547}

\author[0000-0002-9508-3667]{R.~J.~Assef}
\affiliation{N\'{u}cleo de Astronomía de la Facultad de Ingeniería, Universidad Diego Portales, Av. Ej\`{e}rcito 441, Santiago, Chile}

\author[0000-0002-4939-734X]{T.~H.~Jarrett}
\affiliation{Astronomy Department University of Cape Town Private Bag X3 Rondebosch 7701 Republic of South Africa}

\author[0000-0003-0624-3276]{S.~Petty}
\affiliation{NorthWest Research Associates
4118 148th Ave NE
Redmond, WA 98052-5164}

\author{S.~A.~Stanford}
\affiliation{Institute of Geophysics and Planetary Physics, Lawrence Livermore National
Laboratory, Livermore CA 94551}
\affiliation{Department of Physics, University of California, Davis, CA 95616}


\author[0000-0002-9390-9672]{C.-W.~Tsai}
\affiliation{National Astronomical Observatories, Chinese Academy of Sciences, Beijing, 100012, People's Republic of China}
\affiliation{CAS Key Laboratory of FAST, NAOC, Chinese Academy of Sciences, People's Republic of China}
\altaffiliation{Jet Propulsion Laboratory, California Institute of
Technology, 4800 Oak Grove Dr., Pasadena, CA 91109}
\affiliation{Physics and Astronomy Department, University of California, Los Angeles, CA 90095-1547}


\correspondingauthor{S.~E.~Lake}
\email{lake@nao.cas.cn}

\begin{abstract}
The study of the extragalactic background light (EBL) in the optical and near infrared has received a lot of attention in the last decade, especially near a wavelength of $\lambda\approx 3.4\micron$, with remaining tension among different techniques for estimating the background. 
In this paper we present a measurement of the contribution of galaxies to the EBL at $3.4\micron$ that is based on the measurement of the luminosity function (LF) in \cite{Lake:2018b} and the mean spectral energy distribution of galaxies in \cite{Lake:2016}. 
The mean and standard deviation of our most reliable Bayesian posterior chain gives a $3.4\micron$ background of $I_\nu = 9.0\pm0.5 \kJy \sr^{-1}$ ($\nu I_\nu = 8.0\pm0.4 \nW \meter^{-2} \sr^{-1} e\operatorname{-fold}^{-1}$), with systematic uncertainties unlikely to be greater than $2\kJy \sr^{-1}$. 
This result is higher than most previous efforts to measure the contribution of galaxies to the $3.4\micron$ EBL, but is consistent with the upper limits placed by blazars and the most recent direct measurements of the total $3.4\micron$ EBL.
\end{abstract}



\section{Introduction}
The extragalactic background light (EBL) is another name for a fundamental component of the cosmos: the overall spectrum and density of photons in the universe. 
In terms of both quantity of energy and number of photons the dominant component of the EBL is the cosmic microwave background (CMB).\footnote{In some usages the EBL and CMB are regarded as distinct, here we follow the usage in \cite{Cooray:2016} that agrees with the plain meaning of the words.}
While the study of the CMB provides a wealth of information about both the universe at the time those photons were emitted and how the evolving universe has modified those photons, there is a lot to be learned from the study of the EBL frequencies that are dominated by photons emitted at different epochs. 
Studies of the whole of the EBL have revealed that there are, broadly speaking, four peaks in its spectral energy distribution (SED): the CMB in the rough wavelength range $320\micron$ to $22\mm$, a dust emission of galaxies peak from $64$ to $700\micron$, a stellar photospheric peak from $0.15\micron$ to $4\micron$, and an active galactic nuclei (AGN) and stellar remnant peak in X-rays from $4$ to $580\picometer$ \citep[see][]{Cooray:2016}.

While the energy content of the radiation field (of which the EBL is a part) has had a sub-dominant impact on the evolution of space-time, itself, since the redshift of matter-radiation equality \citep[around $z=3000$;][]{Hinshaw:2013}, its spectrum encodes useful information about the history of star and structure formation in the universe. 
With the exception of high energy photons (energy roughly higher than the Lyman-$\alpha$ transition), once the intergalactic medium achieved full reionization the universe became transparent. 
The primary consequence of this is that once a photon escapes from the dense matter in a galaxy halo it has a low probability of ever scattering again, leaving it to eventually be redshifted into oblivion by cosmic expansion. 
This means that when we sample the small fraction of the EBL that reaches our detectors, we are sampling an integrated record of all the light the universe has emitted.

The primary challenge in directly measuring the EBL is removing the large glare of foreground sources, especially the reflected sunlight from dust in our solar system known as the zodiacal light \citep[studies using this approach include: ][]{Gorjian:2000,Wright:2000,Cambresy:2001,Wright:2001,Matsumoto:2005,Levenson:2007,Tsumura:2013,Sano:2016}. 
It is for that reason that two other techniques have come to the fore in the attempts to study the EBL in the optical and near-infrared. 
The first is to study individually detected galaxies, directly, and extrapolating to the undetected galaxies to estimate how much light galaxies have released into the EBL \citep[for example:][]{Fazio:2004,Levenson:2008,Dominguez:2011,Helgason:2012,Driver:2016,Stecker:2016}. 
The challenge with this technique lies in the accuracy of the extrapolation technique used. 
The second is to leverage the fact that low energy photons can pair produce with very high energy photons, providing opacity to them \citep[for example: ][]{Aharonian:2006,Mazin:2007}. 
Ideally this means that high energy gamma rays produced by blazars, typically in the $\operatorname{TeV}$ range of energy, are sampling the EBL directly in intergalactic space. 
The catch that limits the accuracy of this technique is the limit of our ability to determine the gamma ray spectrum pre-extinction, including the production mechanisms and location \citep[see, for example,][]{Essey:2010}.

There is controversy in this field in the measurement of the background in the range of 1--4\micron 
 where direct measures are higher than the upper limits from gamma ray blazars by about a factor of 2, and the lower limits from extrapolating number counts are not definitive. 
With the goal of providing more information relevant to the discussion around 3.4\micron, this work is based on the extrapolation technique. 
Rather than extrapolating the flux histogram, as was done in \cite{Fazio:2004,Levenson:2008,Driver:2016}, this work uses an approach based on extrapolating the galaxy luminosity function (LF), as was done in \cite{Dominguez:2011,Helgason:2012,Stecker:2016}. 
In contrast to previous LF based extrapolation measurements, where they extrapolated extant LFs from the literature, this work is based on a measurement of the LF from scratch that leveraged six public spectroscopic redshift databases of diverse depth and breadth with multiple public photometry databases (especially AllWISE) to construct multiple LFs.
Translating an LF measurement into an EBL estimate requires something equivalent to the mean SED of galaxies, too (previous works used LFs at multiple wavelengths to work around measuring this directly).
To measure this quantity we used only the zCOSMOS data set because of its depth and exceptional variety of photometric information.
Along the way, decisions about what data to use were all optimized for measuring the contribution of galaxies to the EBL at $3.4\micron$. 
The reliance on spectroscopic redshifts reduces the exposure to systematic uncertainties inherent in photometric redshift surveys, and the measurement of a new LF from scratch permits us to feed forward all of the information about the significant correlations among estimated LF parameters into the background estimates.

The structure of this paper is as follows: Section~\ref{sec:theory} covers the equations that relate the LF and mean SED of galaxies to the source flux histogram and integrated background, Section~\ref{sec:datsum} contains a short summary of the results from previous paper used to measure the background and its uncertainty, Section~\ref{sec:res} presents estimates of the $3.4\micron$ background and flux histograms at various wavelengths, and Section~\ref{sec:discussion} discusses how our results compare to previous estimates in the literature.

The cosmology used in this paper is based on the WMAP 9 year $\Lambda$CDM cosmology \citep{Hinshaw:2013}\footnote{\url{http://lambda.gsfc.nasa.gov/product/map/dr5/params/lcdm_wmap9.cfm}}, with flatness imposed, yielding: $\Omega_M = 0.2793,\ \Omega_\Lambda = 1 - \Omega_M$, redshift of recombination $z_{\mathrm{recom}} = 1088.16$, and $H_0 = 70 \km \second^{-1} \Mpc^{-1}$.
All magnitudes will be in the AB magnitude system, unless otherwise specified.
When computing bandpass solar luminosities we utilized the 2000 ASTM Standard Extraterrestrial Spectrum Reference E-490-00\footnote{\url{http://rredc.nrel.gov/solar/spectra/am0/}}. 
For our standard bandpass, W1 at $z=0.38$, we calculate the absolute magnitude of the sun to be $M_{2.4\micron\, \odot} = 5.337\operatorname{AB\ mag}$, $L_{2.4\micron\, \odot} = 3.344 \times 10^{-8} \Jy \Mpc^2$ from that spectrum.

\section{Theoretical Tools}\label{sec:theory}
The basis of the calculations in this work is a mathematical object called the spectro-luminosity functional, denoted $\Psi[L_\nu](\nu)$, that is related to galaxy SEDs in the same way that the ordinary LF is related to regular luminosity.
In words it is the mean number of galaxies per unit comoving volume per unit function space volume.
\cite{Lake:2017} contains a fuller treatment of  $\Psi$. 
One property of $\Psi$ is that the comoving spectral luminosity density $\rho_\nu$ ($L_\nu$ per unit comoving volume, related to the spectral emission coefficient of radiative transfer, $j_\nu$, by a factor of $4\pi$), is the first moment of it:
\begin{align}
	\rho_\nu & = \int [\mathcal{D} L_\nu] L_\nu \Psi[L_\nu](z).
\end{align}
Splitting $\Psi[L_\nu](z)$ into the traditional luminosity function, $\Phi(L_f,z)$, and the likelihood of a galaxy having an SED given that a normalization luminosity is fixed (the normalized SED is $\ell_\nu(\nu) \equiv L_\nu(\nu) / L_f$, with $f= c / [2.4\micron]$, here) gives:
\begin{align}
	\rho_\nu & = \int \d L_f \int [\mathcal{D}\ell_\nu]\, \ell_\nu \Lsed[\ell_\nu](\cdot | L_f, z)\, L_f \Phi(L_f, z) \nonumber \\
	& = \int \d L_f\, \mu_\nu(\nu[1+z],L_f,z)\, L_f \Phi(L_f, z),\label{eqn:rhoL_Lsed}
\end{align}
\noindent where $\Lsed[\ell_\nu](.| L_f, z)$ is the likelihood that a galaxy at redshift $z$ with spectral luminosity $L_f$ at $f = c/[2.4\micron]$ will have the normalized SED given by $\ell_\nu$. 
The dot and vertical pipe character, `$|$', are there to emphasize that $L_f$ and $z$ are treated as non-random variables for \Lsed, in the same way the notation is used for conditional probabilities. 
This is relevant because it affects what units \Lsed\ has; it is a density with respect to the random variables, and not the non-random ones. 
The second line of Equation~\ref{eqn:rhoL_Lsed} is an application of the definition of the mean normalized SED, $\mu_\nu \equiv \langle \ell_\nu \rangle$. 

The transition from spectral luminosity density to the background is achieved by treating each infinitesimal comoving volume element, $\d V_c$, as a galaxy with SED $L_\nu = \rho_\nu \d V_c$ that subtends a solid angle $\d \Omega$. 
The relationship between observed flux surface brightness, $\frac{\d F_\nu}{\d \Omega} \equiv I_\nu$, which is the physical quantity that defines the EBL, then follows from the definition of luminosity distance, $D_L(z)$:
\begin{align}
	F_\nu(\nu_{\mathrm{obs}}) \d \nu_{\mathrm{obs}} & = \frac{L_\nu(\nu_{\mathrm{rest}}) \d \nu_{\mathrm{rest}}}{4 \pi D_L^2(z)} \nonumber \Rightarrow \\
	\frac{ \d I_\nu(\nu)}{\d z} & = \frac{\rho_\nu([1+z]\nu, z)}{4\pi [1+z] D_{\mathrm{cT}}^2(z)} \left(\frac{\d V_c}{\d z} \right) \frac{1}{\Omega_{\mathrm{sky}}} \nonumber \\
	& = \frac{\rho_\nu([1+z]\nu, z)}{\Omega_{\mathrm{sky}} [1+z] } \left(\frac{\d D_c}{\d z} \right), \label{eqn:InuDluma}
\end{align}
where $D_c$ is the radial comoving distance at redshift $z$.

In the absence of emission, cosmological dimming for any surface brightness has four factors of $1+z$ in it, so $I_\nu(\nu,0)\d\nu_{\mathrm{obs}} = (1+z)^{-4} I_\nu(\nu,z)\d\nu_{\mathrm{rest}}$. 
It is useful to use Equation~\ref{eqn:InuDluma} to add the effect of emission to this expression, producing a relation between the EBL at one redshift and the EBL at another.
This allows for the calculation of an evolving EBL, and the ability to combine different models that are valid at different redshifts in a straightforward way. 
That combination is: 
\begin{align}
	I_\nu(\nu, z) & = \frac{[1+z]^3}{\Omega_{\mathrm{sky}}} \int_z^{z_f} \rho_\nu (\nu[1+z'], z')\, c\frac{\d t_L}{\d z'} \d z' \nonumber \\
	&\hphantom{=}+ \left(\frac{1+z_{\hphantom{f}}}{1+z_f}\right)^3 I_\nu(\nu, z_f),
\end{align}
where $c\frac{\d t_L}{\d z} = \frac{1}{1+z} \frac{\d D_c}{\d z}$, and $\rho_\nu$ is defined in Equation~\ref{eqn:rhoL_Lsed}. It should be reiterated that the frequency, $\nu$, is as measured at $z=0$, making it a comoving/coordinate frequency, and $[1+z]\nu$ is the rest frame (physical) frequency at $z$.

Combining Equations~\ref{eqn:rhoL_Lsed} and \ref{eqn:InuDluma} gives the contribution to the background per unit luminosity per unit redshift:
\begin{align}
	\frac{ \d^2 I_\nu}{\d z \d L_\nu} & = \frac{\mu_\nu([1+z]\nu, L_f, z)}{\Omega_{\mathrm{sky}} [1+z]} \left(\frac{\d D_c}{\d z} \right) L_f \Phi(L_f, z). \label{eqn:bgdensity}
\end{align}
Note that the more familiar $K$-correction can be written in terms of $\mu_\nu$ as $K = -2.5\log_{10} \left([1+z]\cdot \mu_\nu \right)$.
The quantity in Equation~\ref{eqn:bgdensity} is useful enough to assign it a symbol of its own, since it is the density of contribution to the background, $\mathcal{B}_\nu(\nu) \equiv \frac{ \d^2 I_\nu}{\d z \d L}$.

Equation~\ref{eqn:bgdensity} can be integrated directly to calculate the predicted background at any frequency, but since the EBL can also be calculated from the integral of the flux times the density of sources per unit flux per unit solid angle,
\begin{align}
	I_\nu & = \int_0^\infty \d F_\nu \, F_\nu\, \frac{\d^2 N}{\d F_\nu \d \Omega},
\end{align}
it is worthwhile to derive an expression for the source counts in order to provide a more detailed check on the LF model in comparison to data not used in the measurement. 
The number of galaxies per unit observed flux per unit solid angle on the sky is given by:
\onecolumngrid 
\begin{align}
	\frac{\d^2 N}{\d F_\nu \d \Omega} & = \frac{1}{\Omega_{\mathrm{sky}}} \int \d z \int \d L_f\, \delta\left(F_\nu - \frac{L_f \mu_\nu(\nu[1+z],L_f,z)}{4\pi[1+z]D_{\mathrm{cT}}^2(z)}\right) \,  \frac{\d V_c}{\d z} \Phi(L_f, z) \nonumber \\
	& = \frac{16\pi^2}{\Omega_{\mathrm{sky}}} \int \d z \frac{\d D_c}{\d z} \left(\frac{[1+z] D_{\mathrm{cT}}^4(z)}{\mu_\nu(\nu[1+z],L_f,z) }\right) \Phi\left(\frac{F_\nu 4\pi [1+z] D_{\mathrm{cT}}^2(z)}{\mu_\nu(\nu[1+z],L_f,z)},\, z\right),
\end{align} 
\twocolumngrid
\noindent where $D_{\mathrm{cT}}(z)$ is the comoving distance transverse to the line of sight (also called $D_M$ for `proper motion' distance).

\section{Summary of Results Used}\label{sec:datsum}
\begin{figure}[htb]
	\begin{center}
	\includegraphics[width=0.48\textwidth]{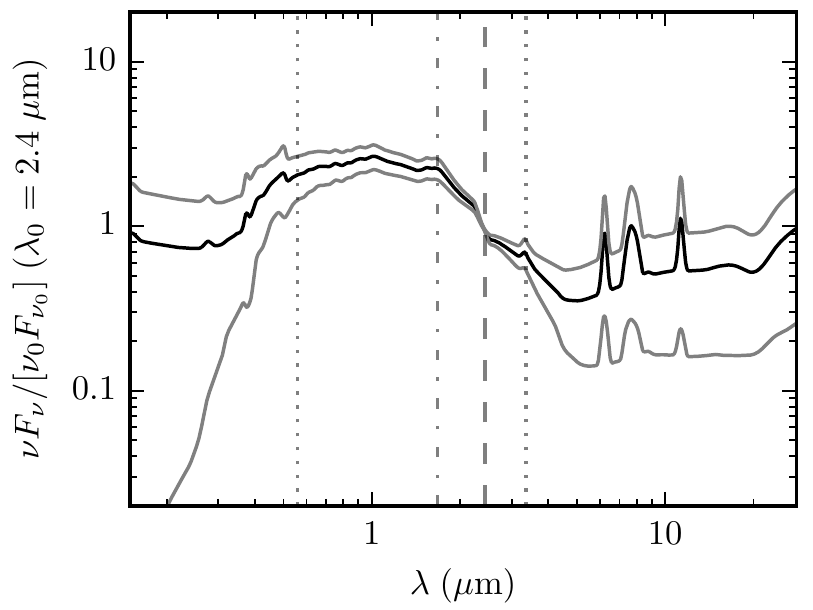}
	\end{center}
	
	\caption{Mean Galaxy SED with Variety Band}{ The mean SED of galaxies as approximated using fits to the templates in \cite{Assef:2010} done in \cite{Lake:2016}. 
	The grey lines show the band of 1-$\sigma$ in SED variety, and the accuracy of the Gaussian approximation is limited to regions where that band is sufficiently narrow compared to the mean SED. 
	The vertical lines are guides to the parts of the mean SED that galaxies at particular redshifts contribute to the $3.4\micron$ background. 
	The vertical dashed line shows the effective rest frame wavelength of W1 for galaxies at $z=0.38$ ($\lambda \approx 2.4\micron$), the vertical dotted lines shows the same for galaxies at $z=0$ and $z=5$, and the vertical dash-dotted line is for galaxies at $z=1$. }
	\label{fig:meanSED}
\end{figure}

As Equation~\ref{eqn:bgdensity} shows, the two necessary quantities for calculating the EBL are the mean SED of galaxies and the LF. 
For the mean SED, this work uses the overall mean from \cite{Lake:2016} that was constructed from fitting templates from \cite{Assef:2010} to targets in the redshift range $z\in(0.05,1]$ (median 0.57) from the zCOSMOS survey described in \cite{Lilly:2009} and \cite{Knobel:2012}. 
The resulting mean SED is shown in Figure~\ref{fig:meanSED}. 
The plot also contains the 1-$\sigma$ band of SED variety around the mean SED because where  the width of that band compared to the position of the mean SED becomes too great determines where the Gaussian approximation of \Lsed\ begins to break down. 
The vertical lines are guides to where particular parts of the mean SED contribute to the $3.4\micron$ EBL.

For the LF this work uses the set of Markov Chain Monte Carlo (MCMC) chains that sample the posterior probability of LF parameters from \cite{Lake:2018b} and available under digital object identifier (DOI) \href{https://dx.doi.org/10.6084/m9.figshare.4109625}{10.6084/m9.figshare.4109625}. 
The measurement that produced the chains is based on a combination of several spectroscopic redshift data sets with public photometric databases, especially the AllWISE data release. 
The LF that the chains have parameters for is a Schechter LF:
\begin{align}
	\Phi(L_f,z) & = \frac{\phi_\star(z)}{L_\star(z)}\left(\frac{L_f}{L_\star(z)}\right)^\alpha e^{-L_f/L_\star(z)}.
\end{align}
The evolution models for $\phi_\star$ and $L_\star$ are similar to the commonly used \cite{Lin:1999}, modified to use lookback time, $t_L(z)$, instead of redshift. 
The model for $L_\star$ also sets $L_\star(t_0) = 0$ at some lookback time $t_0$ when galaxies first lit up, forcing $L_\star$ to peak at some finite redshift.
The exact parameterizations are:
\begin{align}
	\phi_\star(t_L) & = \phi_0 e^{-R_\phi t_L},\ \mathrm{and} \\
	L_\star(t_L) & = L_0 e^{-R_L t_L} \left(1 - \frac{t_L}{t_0}\right)^{n_0},
\end{align}
where $\phi_0$, $R_\phi$, $L_0$, $R_L$, $\alpha$, $n_0$, and $t_0$ are all constants. 
The only constant not, in some way, measured is $t_0$ which is set to the lookback time of recombination, equivalent to the redshift $z_{\mathrm{recomb}} = 1088.16$ according to the WMAP 9 year $\Lambda$CDM parameters matrix of \cite{Hinshaw:2013}\footnote{\url{https://lambda.gsfc.nasa.gov/product/map/current/params/lcdm_wmap9.cfm}} (giving $t_0 = 0.9828\, t_H = 13.73 \Gyr$).
Using $z_{\mathrm{reion}}=10.6$ gives similar results, despite the fact that galaxies must have been producing light before then, so the results are not very sensitive to the details of $t_0$, as long as $L_\star(t_0)$ is small enough to force a turnover in $L_\star(t)$.
Full details of how to extract these parameters from the contents of the chain files are given in \cite{Lake:2018b}.

\cite{Lake:2018b} has 12 different MCMC posterior chains: one for each of the six surveys, and six that combine the data in different ways (primarily to work around an issue with bright sources in the low redshift data).
Producing this many different analyses gives a good handle on any systematic issues.
Of the 12 chains, the two combined data ones that use high redshift data with a prior on $\alpha$ are preferred because the data from many redshifts is needed to constrain the evolution rate parameters, to which the background is very sensitive. 
These samples are denoted ``High $z$ Prior" and ``High $z$ Trim Prior".
Of the two, the former is preferred because it has a much larger sample size, though the latter is less sensitive to the details of the spectro-luminosity model that, effectively, determined the completeness model.

For every set of parameters in the LF chains two backgrounds are calculated: the background from sources with $z \le 1$, and from sources with $z\le 5$. 
The first redshift limit is set to match the upper limit on the data used to measure the LFs, and so marks the lower edge of model validity. 
The upper limit at $z=5$ is set to capture as much of the background predicted by the model as possible without relying on the parts of the mean SED where the size of the SED covariance makes the Gaussian approximation of \Lsed\ invalid.

\section{Backgrounds and Number Counts}\label{sec:res}
The primary result from \cite{Lake:2018b} is a collection of posterior MCMC chains for evolving luminosity function parameters. When combined with the mean SEDs from \cite{Lake:2016}, it is possible to calculate a couple of different interesting quantities. First, we combine each element from the chains to calculate a background observable now, giving MCMC posterior chains for the EBL. Second, we use the mean LF parameters from the `High $z$ Prior' chain to calculate predicted histograms of galaxies, and compare them to observed histograms of all sources. Third, using the same combination used to calculate the number counts, we also calculate an evolving spectral luminosity density and EBL as a function of comoving wavelength from $0.5$--$5\micron$, quantities useful in predicting the total opacity to very high energy gamma rays.

\begin{figure*}[htb]
	\begin{center}
	\includegraphics[width=\textwidth]{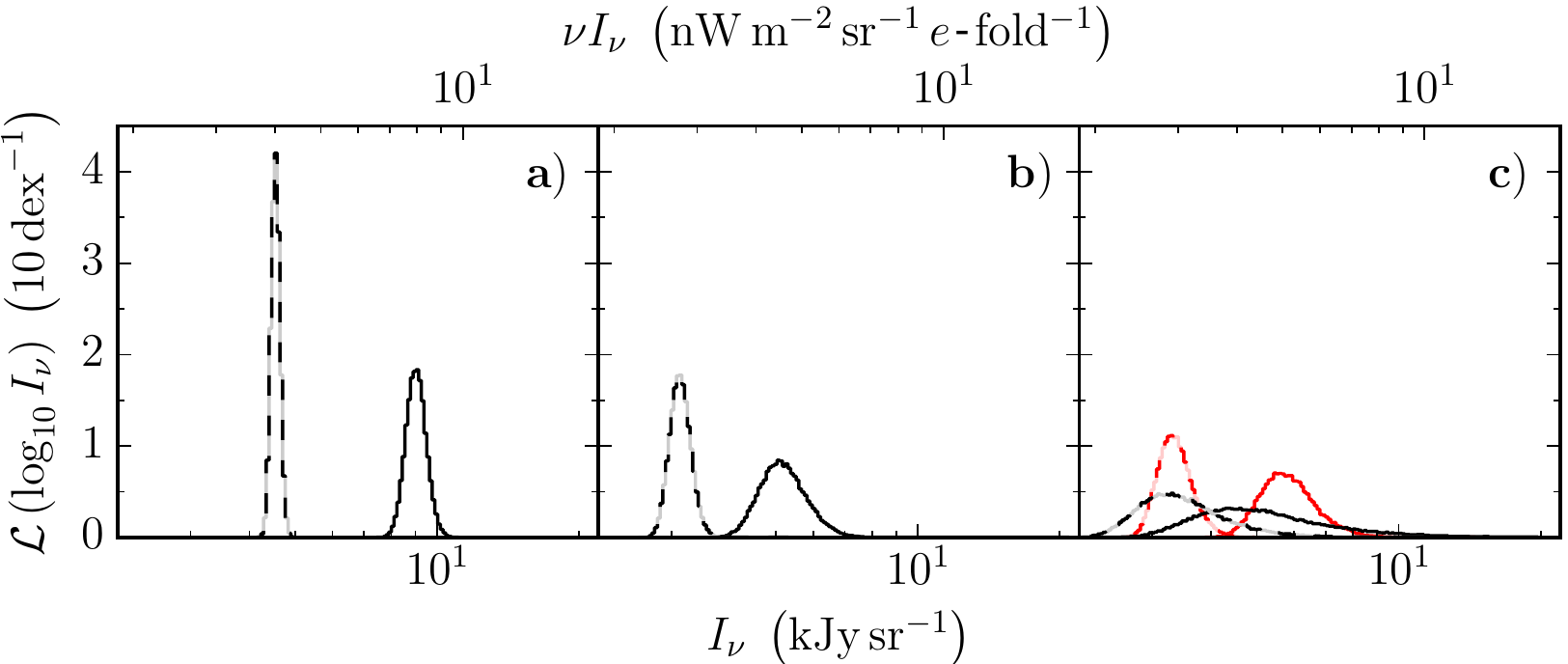}
	\end{center}
	
	\caption{Example EBL Histograms}{ Histograms of EBL posteriors that correspond to different LF posteriors from \cite{Lake:2018b}. 
	The dashed lines are histograms of the $z\le 1$ predictions and the solid lines are of the $z\le 5$. 
	Panel~\textbf{a} is based on the High $z$ Prior chain, Panel~\textbf{b} is based on the High $z$ Trim Prior chain, and Panel~\textbf{c} uses the survey specific chains from \WD\ (black) and zCOSMOS (red). 
	Each posterior chain contains a total of $210,000$ samples. }
	\label{fig:BGhists}
\end{figure*}

The backgrounds that correspond to the posterior MCMC chains from \cite{Lake:2018b} are published alongside this article as a machine readable table, and under \href{https://dx.doi.org/10.6084/m9.figshare.4245443}{DOI:10.6084/m9.figshare.4245443}. 
Each posterior LF chain has a corresponding file containing the backgrounds calculated for each element in the LF chain. 
Examples of histograms constructed using the posterior chains can be found in Figure~\ref{fig:BGhists}, and a few example lines from one of the posterior tables are in Table~\ref{tbl:BGposterior}.
Except for the smallest data set plotted, the black lines of Panel~\textbf{c}, all of the distributions are visually similar to log-normal distributions. 

\begin{deluxetable}{rrcc}
	\tabletypesize{\scriptsize}
	\tablewidth{0.45\textwidth}
	\setlength{\tabcolsep}{5.0pt}
	\tablecaption{Example Lines from MCMC Background Chains}
	\tablehead{  \colhead{\code{StepNum}} & \colhead{\code{WalkerNum}} &
			\colhead{\code{Inu\_1}} & \colhead{\code{Inu\_5}} \\
		\colhead{---} & \colhead{---}  &
			\colhead{$\kJy \sr^{-1}$}  & \colhead{$\kJy \sr^{-1}$}  }
	\startdata
		    $0$     &   $0$  &     $4.6784$   &   $9.8475$ \\    
		    $0$     &   $1$  &     $4.6598$   &   $9.3267$ \\    
		    $0$     &   $2$  &     $4.4538$   &   $8.7004$ \\   
		    $0$     &   $3$  &     $4.5978$   &   $9.2741$ \\    
		    $0$     &   $4$  &     $4.5714$   &   $9.1531$ \\
	\enddata
	\tablecomments{Example lines from one of the chains produced by \code{emcee} in the tables under DOI \href{https://dx.doi.org/10.6084/m9.figshare.4245443}{\code{10.6084/m9.figshare.4245443}}. 
	Floating point values truncated here for brevity, but not in the downloadable tables. 
	\code{StepNum} is the zero indexed step number that the ensemble was at in the chain, and \code{WalkerNum} is the number of the walker which was at the position defined by the row for that step. 
	\code{Inu\_1} is the contribution of galaxies to the EBL at $3.4\micron$ for the corresponding element of the chain from \cite{Lake:2018b} for galaxies at redshift $z<1$.
	\code{Inu\_5} is the same but for galaxies with redshift $z<5$. }
	\label{tbl:BGposterior}
\end{deluxetable}

\begin{figure*}[htb]
	\begin{center}
	\includegraphics[width=\textwidth]{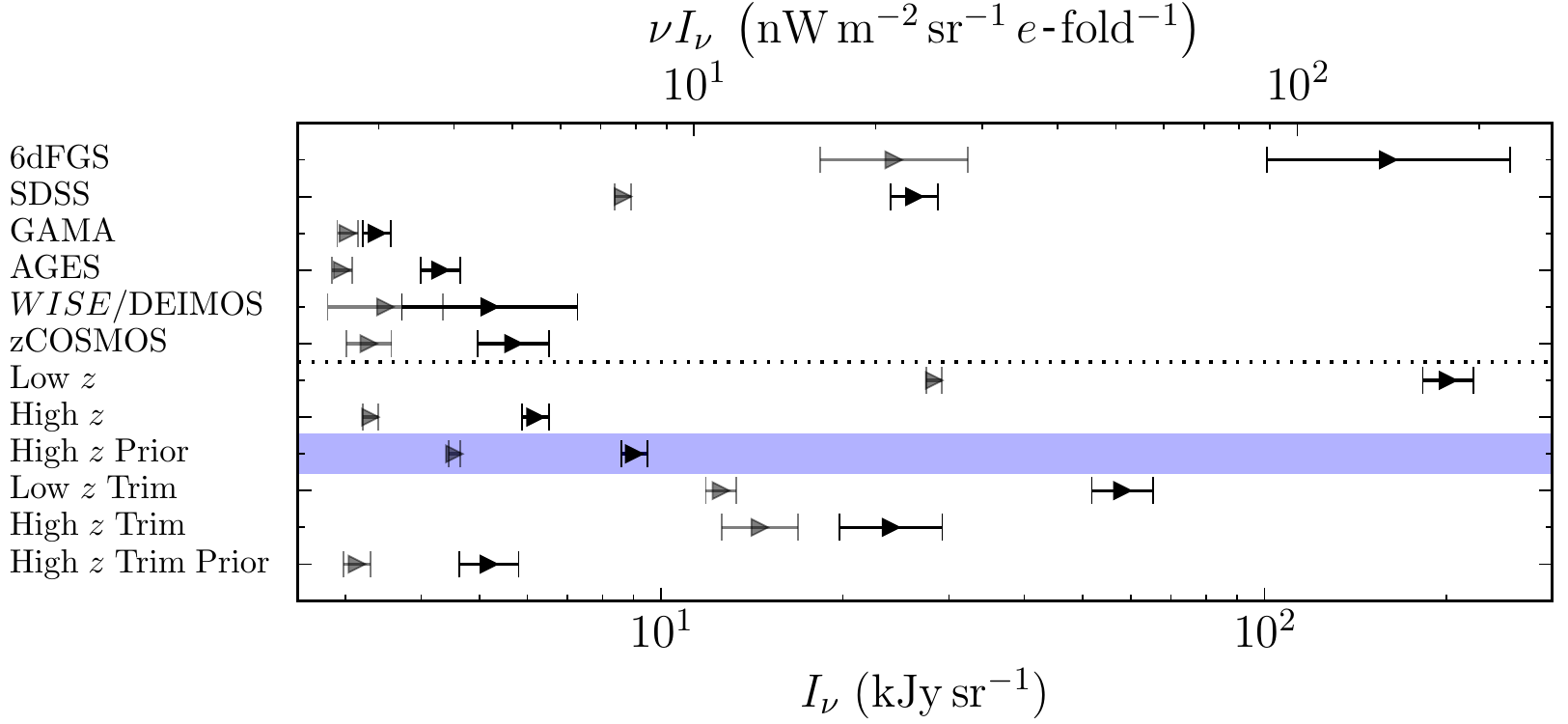}
	\end{center}
	
	\caption{EBL and Uncertainties for Different Chains}{ All of the $3.4\micron$ EBL predictions made using the posterior chains from \cite{Lake:2018b}. 
	The grey points are the $z\le 1$ backgrounds, and the black points are the $z\le 5$ ones. 
	The dotted lines divides the survey based samples (above) from the combined samples (below). 
	The blue bar highlights background based on the canonical chain from \cite{Lake:2018b}.  }
	\label{fig:EBLbychain}
\end{figure*}

The symmetry of the histograms in log-space makes a description of each result as a log-normal distribution, parameterized by its geometric mean, $\langle \log_{10} I_\nu \rangle$, and logarithm standard deviation, $\left\langle \left(\log_{10} I_\nu - \langle \log_{10} I_\nu \rangle\right)^2\right\rangle$, a good approximation of the whole distribution. 
The means and standard deviations for all of the log-background posterior chains can be found in Figure~\ref{fig:EBLbychain}, with the blue bar highlighting the official result for this paper, $I_\nu(\lambda = 3.4\micron) = 9.03^{+0.46}_{-0.43} \kJy \sr^{-1}$ ($\nu I_\nu = 7.96^{+0.40}_{-0.38} \nW \meter^{-2} \sr^{-1} e\text{-fold}^{-1}$). 
The full details of what defines all of the samples that fix the model parameters can be found in \cite{Lake:2018b}. 
The survey specific samples (above the dotted line) are sorted in order of increasing depth (defined as median redshift) with shallowest on top. 
The combined samples (below the dotted line) are Low $z$ when the data is limited to redshift $z \le 0.2$, High $z$ when $0.2 < z \le 1.0$, Prior when the faint end slope ($\alpha$) of the LF is constrained using the mean and standard deviation of the faint end slope of the corresponding Low $z$ sample, and Trim when the contributions of each survey are limited to areas in the luminosity-redshift plane where the survey is more than 98\% of its maximum completeness. 
In sum, the Trim samples sacrifice sample size and depth for reduced systematic uncertainty, and the Prior samples combine the aspect of the Low $z$ samples least affected by a bias of uncertain origin that affects bright sources. 

The important features to note in Figure~\ref{fig:EBLbychain} are: the samples most affected by the unknown bias identified in \cite{Lake:2018b} (6dFGS, SDSS, Low $z$, and Low $z$ Trim) have the expected backgrounds so high they can accurately be described as outliers, and there is an increasing trend in the predicted background with survey depth (GAMA, AGES, \WD, and zCOSMOS, in order). 
The presence of the High $z$ Trim sample in the category of outliers is a consequence of that chain having a mean faint end slope of $\alpha = -1.93\pm -0.04$, nearer the point where the LF estimate diverges at $\alpha=-2$ than the other samples which all have $\alpha$ nearer to $-1$. 
It is likely that the same fluctuation that makes the $\alpha$ of the High $z$ Trim sample so negative was displaced into a faster comoving number density evolution; one that implies galaxies are presently increasing in comoving number density at $1.9\pm0.7$ $e$-folds per Hubble time, reducing the estimated background. 
For all of these reasons, and because it has the greatest statistical precision, the High $z$ Prior is most likely to prove most accurate when compared with even better, deeper, measurements made in the future.

\begin{figure}[htb]
	\begin{center}
	\includegraphics[width=0.48\textwidth]{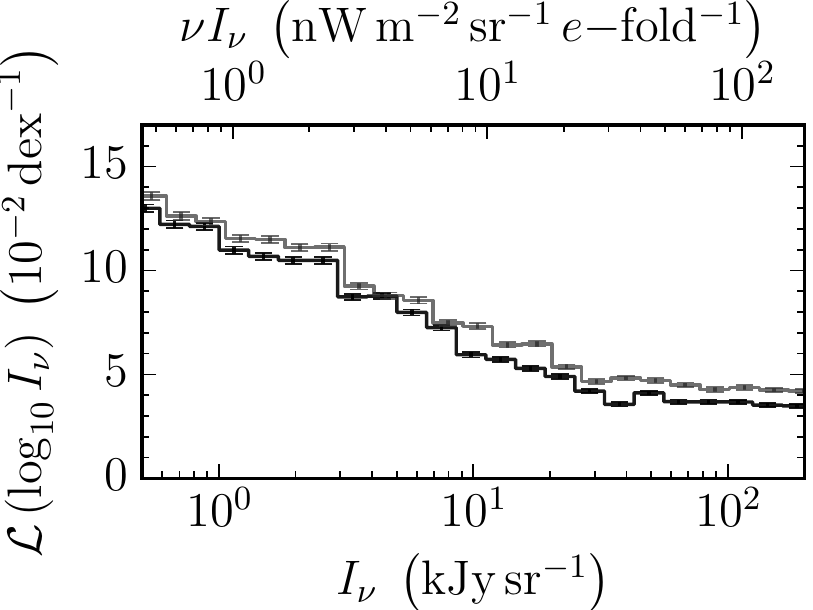}
	\end{center}
	
	\caption{Induced Prior on the EBL Estimates}{ The relevant part of the EBL prior induced by the priors on the LF parameters in \cite{Lake:2018b}.
	The black line is the prior for $z\le 5$ backgrounds, and the grey line is for the $z\le 1$ backgrounds. 
	The grey line is shifted to the right by $26\,$milli-dex. 
	The chain used to construct this histogram contained $322,560$ samples, in total, spanning more than 30 orders of magnitude (the priors on the LF parameters were very broad), and the error bars are approximated by assuming Poisson statistics. 
	Note that the black and grey lines are not independent, since they were calculated using the same MCMC LF parameter chains. }
	\label{fig:EBLprior}
\end{figure}

The priors imposed on the LF parameters in \cite{Lake:2018b} were chosen to be analytically calculable and minimally informative, with the exception of the parameter that defines the poorly constrained early time behavior of $L_\star$, $n_0$. 
Just as MCMC permits drawing a set of samples from the posterior distribution, it is also possible to sample the prior. 
This is not usually done, because priors are usually analytically calculable. 
In this case, though, the prior on the EBL values is not analytically calculable from the prior on the LF parameters (primarily due to the numerical definition of the mean galaxy SED).
The EBL prior must, therefore, be reconstructed from Monte Carlo samples drawn from the LF prior.
Because the EBL prior is entirely determined priors set on other parameters, it is described here as being `induced' implicitly by the prior on the LF parameters.
A portion of a histogram of the induced EBL prior chain can be found in Figure~\ref{fig:EBLprior} (the range was restricted to what is relevant for the posterior chains in this work). 
Some example lines from the EBL prior chain file can be found in Table~\ref{tbl:EBLprior}. 
The black line is the histogram of the $z\le 5$ background, and the grey line is of the $z\le 1$ background, offset to the right by $26\,$milli-dex, for clarity. 
The prior is clearly not flat in either $I_\nu$ or $\ln I_\nu$ in the region of interest. 
It is, in fact, slightly biased to the low side. 
Note that the response of a log-normal distribution, $f(\ln(x)) \propto \exp(- [\ln x - \mu]^2 / [2 \sigma^2] )$, to a prior that is approximable as $x^k$ (here $k\approx -0.5$ for most of the range of interest) is to shift it by an amount that depends on the width of the distribution, $\delta \mu = k \sigma^2$. 
Because $|k\sigma| < 1$ for all of the posteriors produced here, the shift will be less than $\sigma$ in all cases; for the High $z$ Prior result, in particular, the shift in the mean caused by the prior is the same as dividing the mean $I_\nu$ by $0.998$.
The backgrounds that correspond to the induced EBL prior MCMC chain are published alongside this article as a machine readable table, and under \href{https://dx.doi.org/10.6084/m9.figshare.8142284}{DOI:10.6084/m9.figshare.8142284}. 

\begin{figure*}[p]
	\begin{center}
	\includegraphics[width=\textwidth]{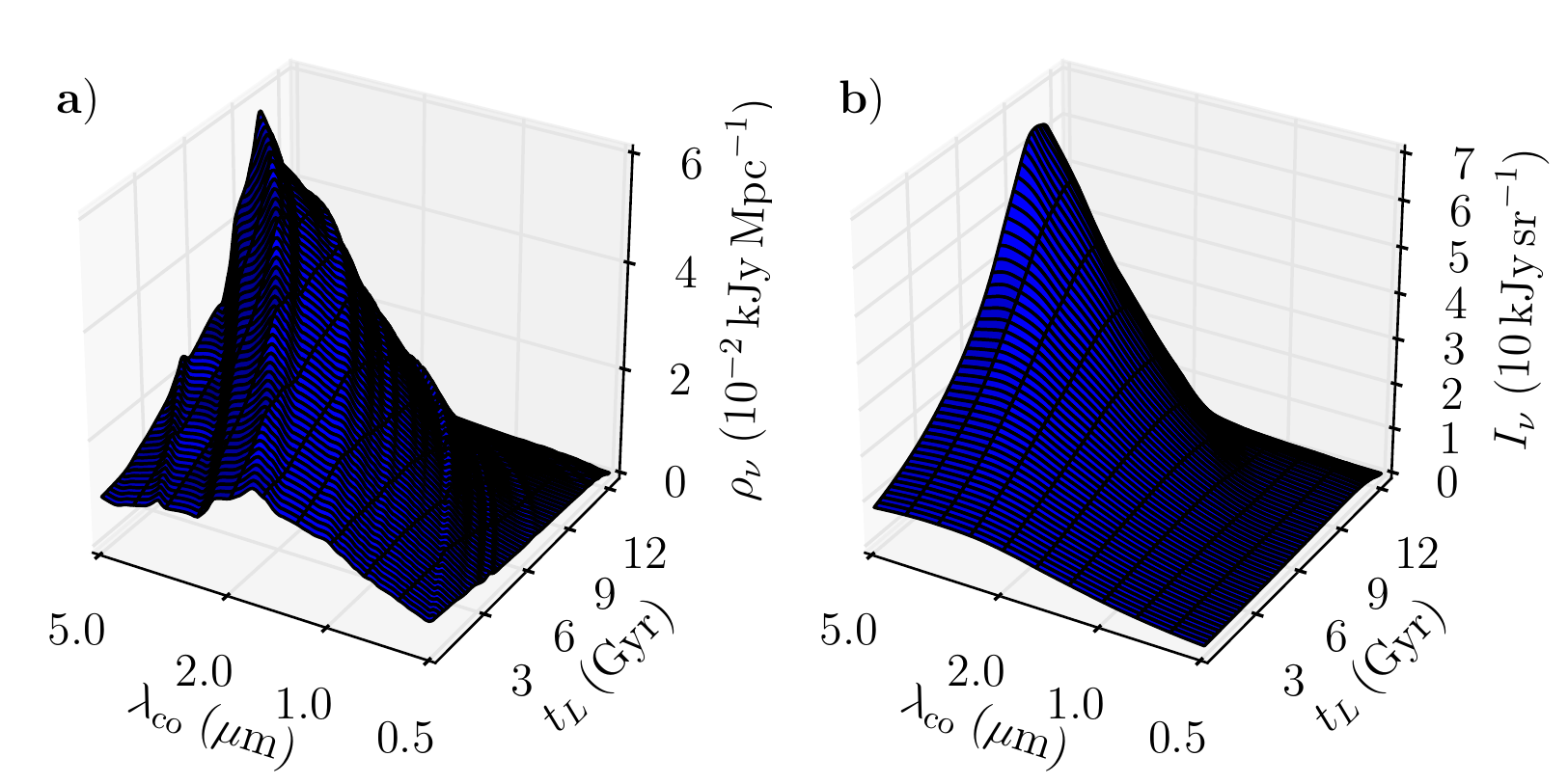}\\
	\includegraphics[width=\textwidth]{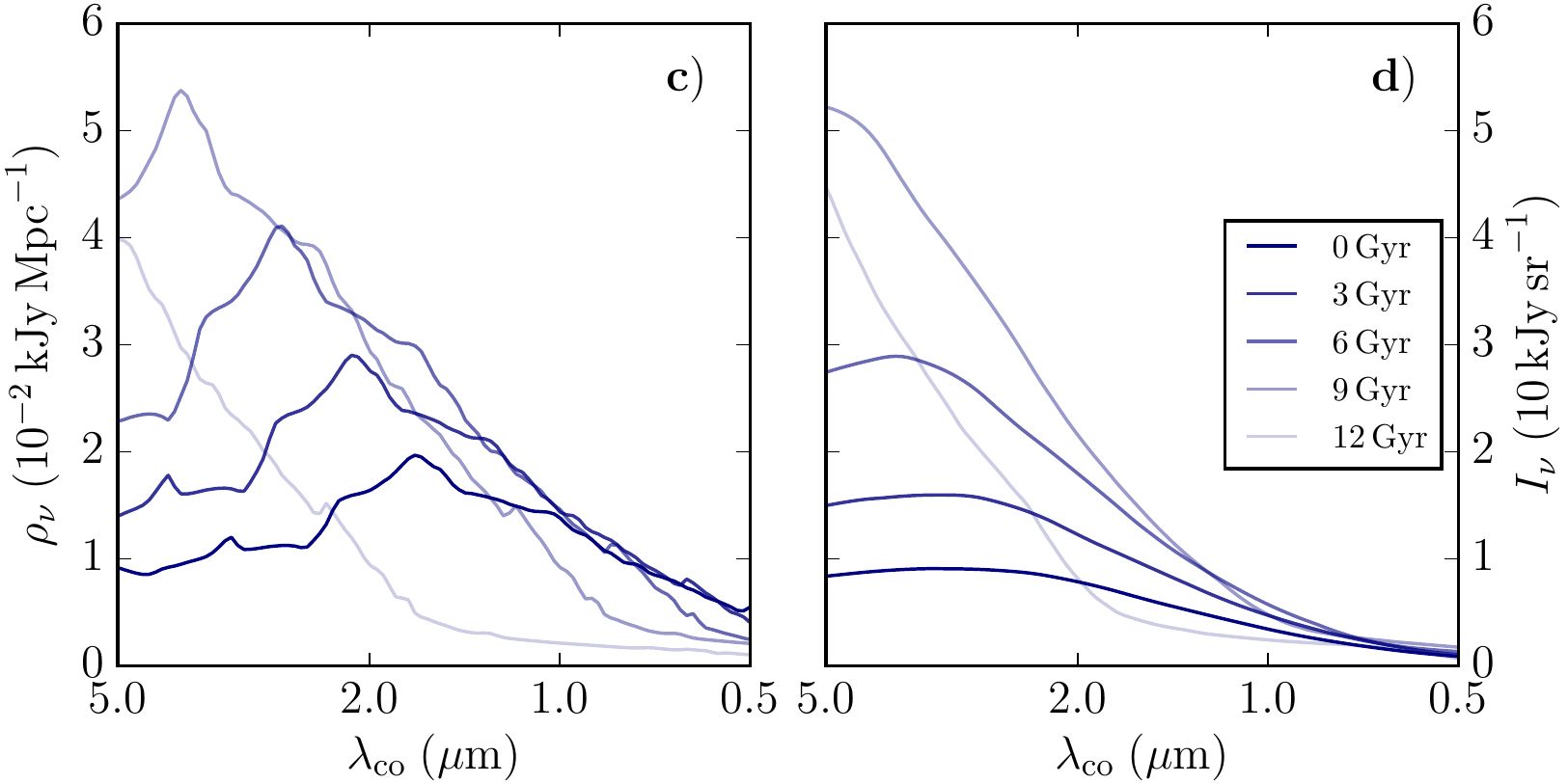}
	\end{center}
	
	\caption{Evolving Luminosity Density and Background}{Panels \textbf{a} and \textbf{b} are 3-dimensional plots of the comoving spectral luminosity density (see Equation~\ref{eqn:rhoL_Lsed}) and the EBL ($I_\nu$) as a function of comoving/coordinate wavelength ($\lambda_{\mathrm{co}}$) and lookback time ($t_L(z)$). The redshifts spanned are from $0$ to $5$. Panels \textbf{c} and \textbf{d} are constant time slices from the above panels (at $0$ through $12 \Gyr$ in steps of $3\Gyr$), with decreasing opacity (darkness) as $t_L$ increases. Note how the EBL roughly tracks the luminosity density at the given epoch, though the detailed spectral features are smeared out. }
	\label{fig:EvolvEBL}
\end{figure*}

The evolving spectral luminosity density and EBL are plotted in Figure~\ref{fig:EvolvEBL}. 
The evolution in the luminosity density, depicted in panels \textbf{a} and \textbf{c}, arises entirely from normalization (evolution of luminosity function parameters) and redshifting of the mean SED. 
The evolving EBL, depicted in panels \textbf{b} and \textbf{d}, loses the spectral features visible in the luminosity density, as the integration smears them out and expansion redshifts old photons away. 
Most other works depict the optical region of the EBL peaking between $1$ and $2\micron$ (see, for example, \cite{Dominguez:2011} and \cite{Cooray:2016}), but that  is the result of plotting $\nu I_\nu$, where Figure~\ref{fig:EvolvEBL} plots $I_\nu$. 
The data used to produce the plots are included with this work as fits tables, and under \href{https://dx.doi.org/10.6084/m9.figshare.4757131}{10.6084/m9.figshare.4757131}. 

%

\begin{deluxetable}{cc}
	\tablewidth{0.45\textwidth}
	\tablecaption{Example Lines from MCMC Induced Prior Background Chains}
	\tablehead{  \colhead{\code{Inu\_1}} & \colhead{\code{Inu\_5}} \\
			\colhead{$\kJy \sr^{-1}$}  & \colhead{$\kJy \sr^{-1}$}  }
	\startdata
		       $2.0879$  &   $2.0879$ \\  
		       $0.5818$  &   $0.5818$ \\   
		       $0.5818$  &   $0.5818$ \\   
		       $0.5818$  &   $0.5818$ \\   
		       $0.5818$  &   $0.5818$  \\
	\enddata
	\tablecomments{Example lines from the induced background posterior chain produced by \code{emcee} in the table under DOI \href{https://dx.doi.org/10.6084/m9.figshare.8142284}{\code{10.6084/m9.figshare.8142284}}. 
	Floating point values truncated here for brevity, but not in the downloadable tables. 
	\code{Inu\_1} is the contribution of galaxies to the EBL at $3.4\micron$ for galaxies at redshift $z<1$,  and \code{Inu\_5} is the same but for galaxies with redshift $z<5$. }
	\label{tbl:EBLprior}
\end{deluxetable}

\section{Discussion}\label{sec:discussion}
\begin{figure*}[htb]
	\begin{center}
	\includegraphics[width=\textwidth]{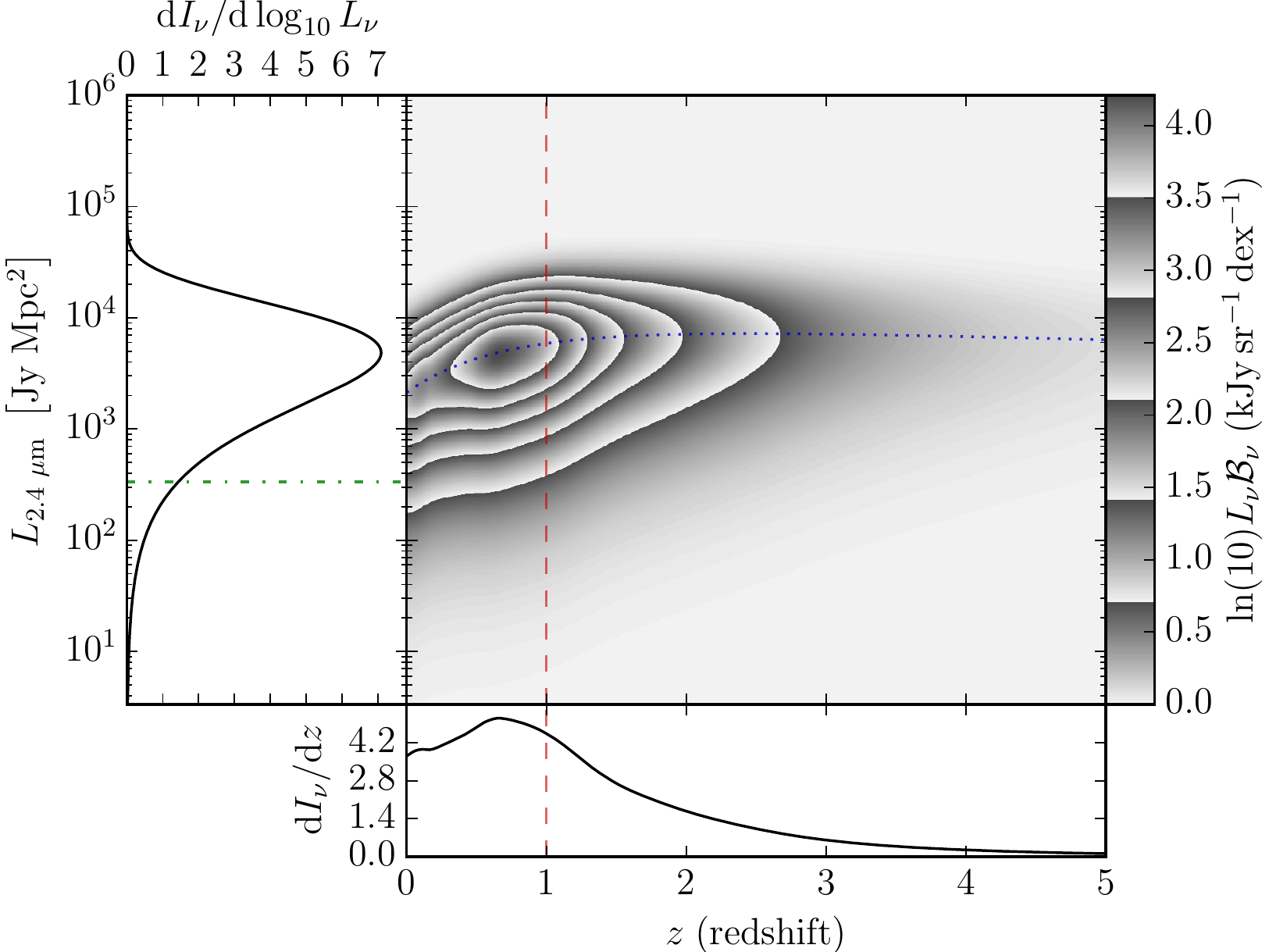}
	\end{center}
	
	\caption{Density of Contributions to the EBL by Redshift and Luminosity}{ Bivariate density of contributions to the $3.4\micron$ EBL by galaxies according to the mean model from the High $z$ Prior MCMC posterior chain, with the marginal densities abutting. 
	The blue dotted line on the bivariate density shows the evolution of $L_\star$, the red vertical dashed line highlights the extent of the data the models were fit to, and the green dash-dotted line marks $10^{10} L_{2.4\micron \odot}$ (where the galaxy's spectral luminosity, $L_\nu$, at wavelength $\lambda = 2.4\micron$ is the same as $L_\nu$ at the same point in the spectrum as $10^{10}$ suns). 
	The total $3.4\micron$ background in this plot is $9.06 \kJy \sr^{-1}$ ($7.99 \nW \meter^{-2} \sr^{-1} e\text{-fold}$). }
	\label{fig:EBLdensity}
\end{figure*}

The models used to calculate the $3.4\micron$ EBL in this work have two shortcomings most relevant to the EBL estimate. 
First, the mean SED did not evolve with luminosity or redshift. 
Second, the faint end slope of the LF was also a constant. 
Figure~\ref{fig:EBLdensity} contains a plot highlighting where in redshift-luminosity space the $3.4\micron$ EBL originates. 
As expected, the model predicts that the EBL is dominated by low to moderate redshift objects with luminosities near $L_\star$. 
Studies of the high redshift universe consistently show that galaxies in the early universe have optical and ultraviolet spectra dominated by high mass stars, making them, on average, more blue than low redshift galaxies. 
\cite{Bouwens:2009} and \cite{Bouwens:2014}, for example, found that galaxies exhibited a trend of bluer ultraviolet slope, decreasing $\beta = \frac{\d \ln L_\lambda}{\d \ln \lambda}$, for both increasing redshift and decreasing luminosity. 
Similarly, studies of the mass function of galaxies show a trend of decreasing $\alpha$ with increasing redshift \citep[see the compilation in Table~1 of][]{Conselice:2016} to a minimum around the $-2$ predicted by the $\Lambda$CDM models in \cite{Jenkins:2001}.

Both color evolution and faint end slope evolution suggest that the EBL measurements produced here are underestimates. 
The former would tend to increase size of the high redshift tail of the bottom plot in Figure~\ref{fig:EBLdensity}, and the latter would increase the low luminosity tail of the left hand plot in the same figure. 
These factors are counter-balanced by the fact that the model has slower evolution in $L_\star$ at high redshift than would be suggested by comparisons with Figure~9 of \cite{Madau:2014}, which would narrow the high redshift tail. 

It is difficult to predict how these competing factors will work out when more accurate measurements are available. 
The SED evolution is unlikely to contribute more than a factor of $10$ to the thickness of the high redshift tail. 
Evolution in $L_\star$ is likely to be of a similar size. 
The interesting challenge is presented by the evolution in $\alpha$, where measurements with $\alpha \le -2$ require the explicit addition of a low luminosity cutoff to the LF to produce a finite contribution to the background. 
This means that fully constraining the contribution of galaxies to the EBL will require measuring galaxies on the faint end slope of the LF to either eliminate $\alpha \le -2$ or to find the LF's faint end cutoff. 
Interestingly, the presence of a faint end cutoff, $L_{\mathrm{min}}$, with a steep faint end slope, $\alpha \le -2$, increases the impact that $L_\star$ evolution has on the predicted EBL because $L_{\mathrm{min}}$ enters into EBL calculations in a ratio with $L_\star$.

\begin{figure*}[htb]
	\begin{center}
	\includegraphics[width=\textwidth]{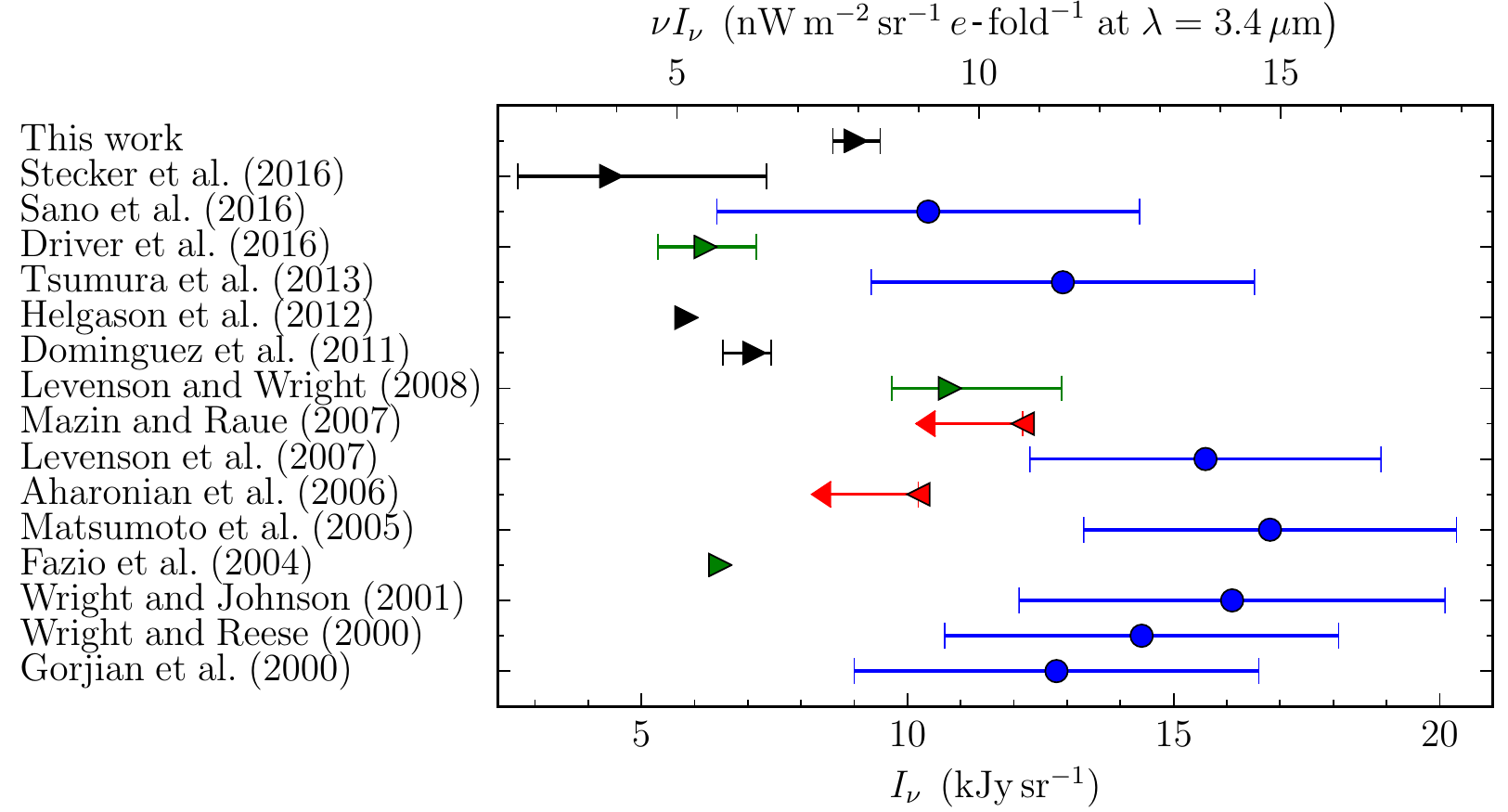}
	\end{center}
	
	\caption{$3.4 \operatorname{\mu m}$ EBL Measurements}
	{Comparison of different values measured for the EBL for wavelengths near $3.4\micron$ ordered by year of publication then first author last name. 
	The points plotted as blue circles are direct observations of the EBL, the green triangles are based on integrating extrapolated galaxy flux histograms to $0$, the red inverted triangles are upper limits from blazar extinction models, and the black triangles are luminosity function based estimates of what galaxies contribute to the EBL. 
	Points without error bars \citep{Fazio:2004, Helgason:2012} did not include error estimates in the original work. }
	\label{fig:externalBGcomp}
\end{figure*}

Figure~\ref{fig:externalBGcomp} shows how the primary estimate in this work compares with values from the literature from wavelengths in the range $3.4$ to $3.6\micron$, adjusted to $3.4\micron$ assuming $I_\nu$ is approximately a constant with wavelength. 
Points in Figure~\ref{fig:externalBGcomp} include direct observations of the EBL \citep{Sano:2016,Tsumura:2013,Levenson:2007,Matsumoto:2005,Wright:2001,Wright:2000,Gorjian:2000}, upper limits based on the examination of the extinction of $\TeV$ gamma rays from blazar spectra \citep{Mazin:2007,Aharonian:2006}, estimates of the contribution of galaxies based on extrapolating galaxy source flux counts \citep{Driver:2016,Levenson:2008,Fazio:2004}, and other LF based estimates \citep{Stecker:2016,Helgason:2012,Dominguez:2011}. 
What can be seen from the comparisons is that, while the $3.4\micron$ EBL measured here is higher than measurements from most comparable works, it is still consistent \textbf{with} the blazar limits and the direct measurements. 
Judging by the relationship to the literature measurements, any modification from the true value caused by the systematic limitations in this work is unlikely be more than about $2\kJy \sr^{-1}$ ($1.8 \nW \meter^{-2} \sr^{-1} e\text{-fold}$) in either direction.

\begin{figure*}[htb]
	\begin{center}
	\includegraphics[width=\textwidth]{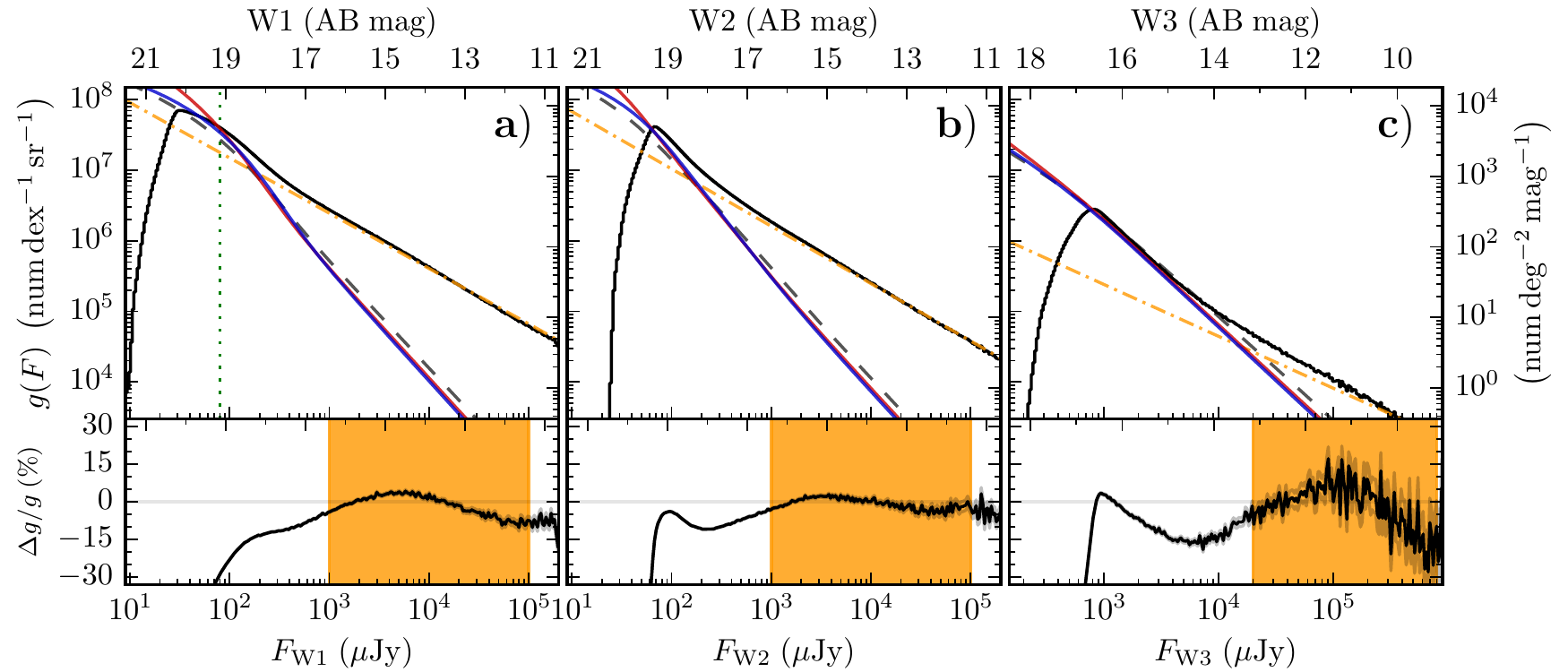} 
	\end{center}
	
	\caption{WISE Flux Counts Comparisons}
	{Comparisons of the predicted source flux counts, $g(f) \equiv \frac{\d^2 N}{\d \log_{10} F \d \Omega}$ from the MCMC chain mean models of \cite{Lake:2018b}, to observed flux counts (solid black line) in the AllWISE W1, W2, and W3 bands in the top panels.
	The bottom panels show the residuals of the data with respect to the High $z$ Prior model added to a power law `star' model. 
	The three galaxy predictions plotted are based on the mean LFs from the samples \cite{Lake:2018b} labeled as High $z$ Prior (red line), High $z$ Trim Prior (blue line), and \WD\ (grey dashed line). 
	 A power law model (orange dash-dotted line) was added to the High $z$ Prior model and fit to the data in the region highlighted in orange in each plot's lower panel.}
	\label{fig:WNumCounts}
\end{figure*}

\begin{figure*}[htb]
	\begin{center}
	\includegraphics[width=\textwidth]{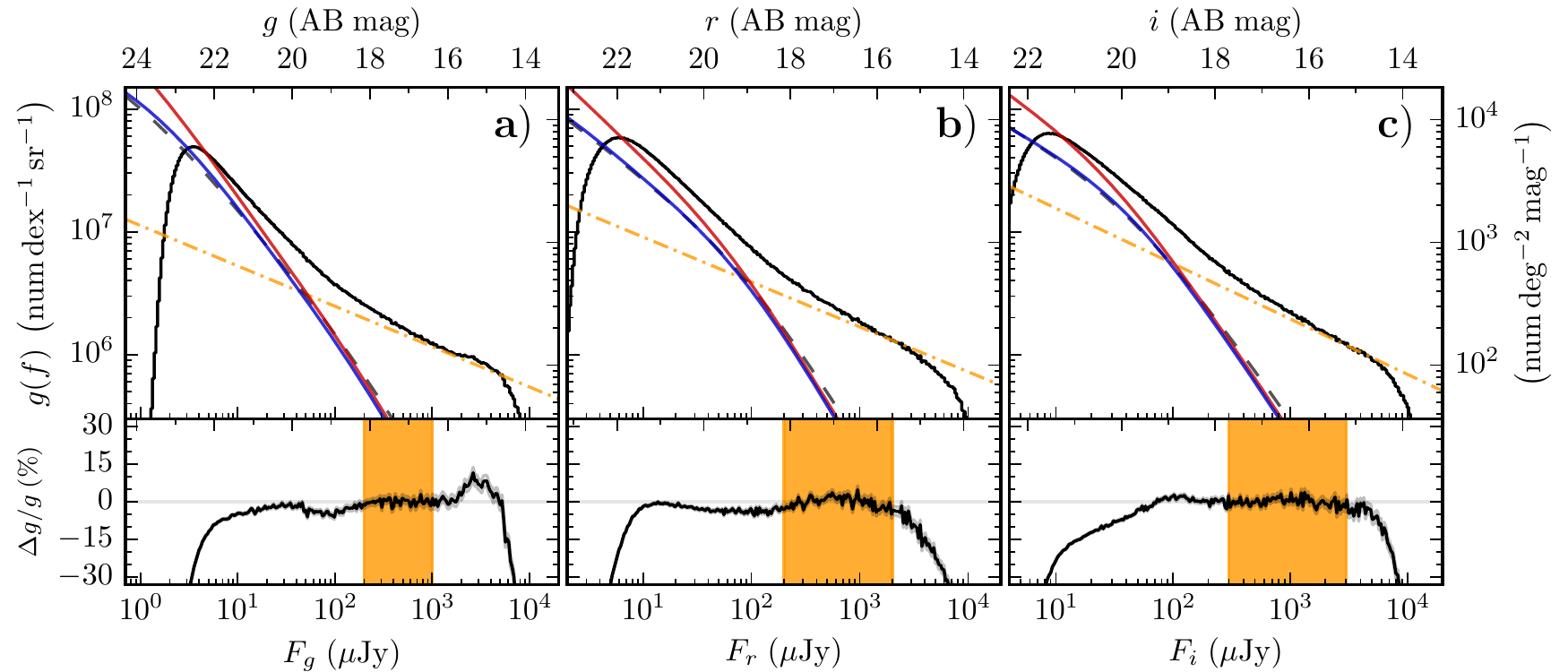} \\
	\includegraphics[width=0.48\textwidth]{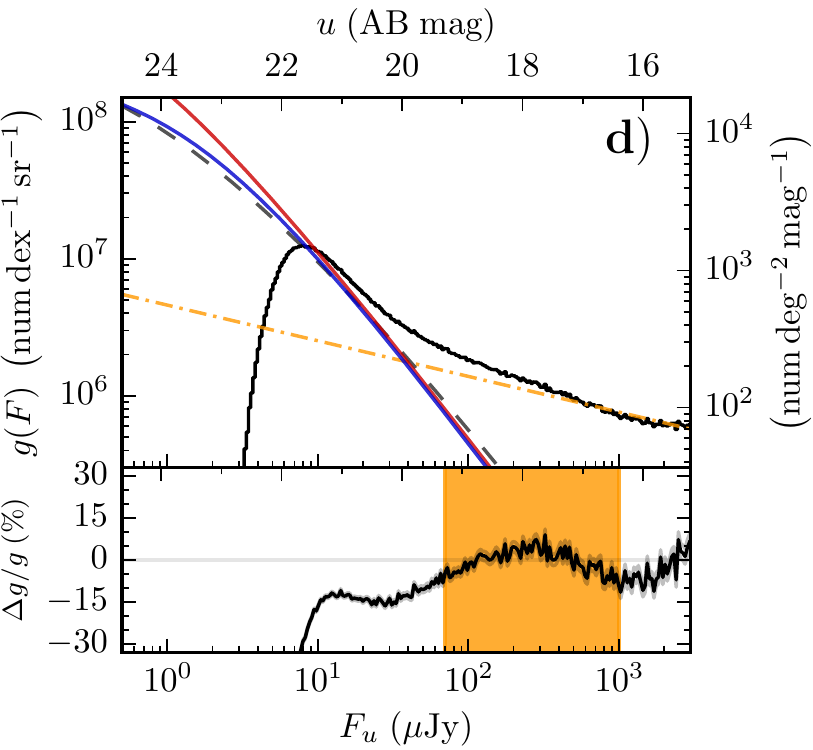}
	\includegraphics[width=0.48\textwidth]{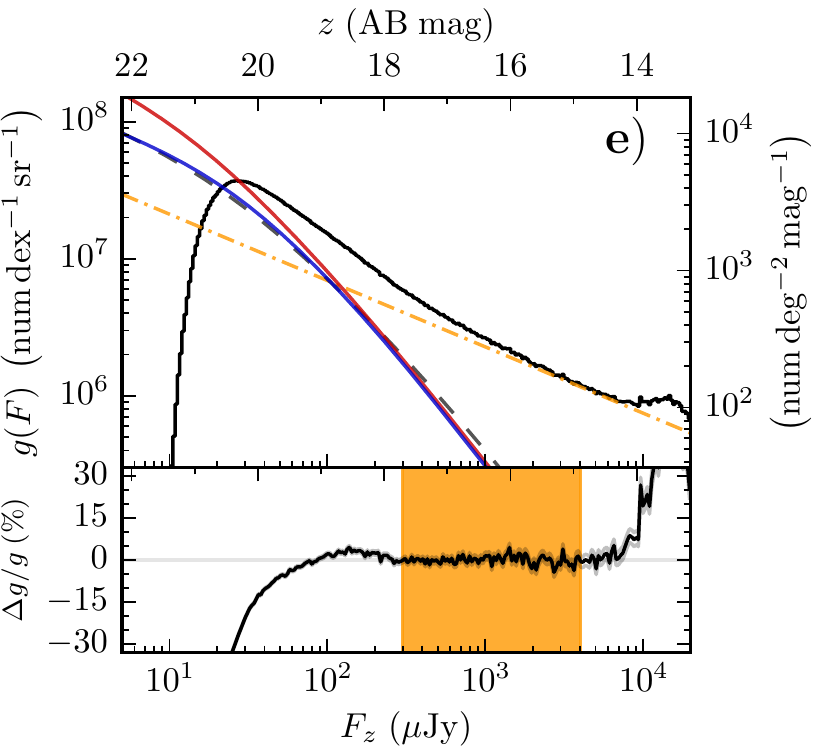}
	\end{center}
	
	\caption{SDSS Flux Counts Comparisons}
	{The same comparison as shown in Figure~\ref{fig:WNumCounts}, but to a sample of SDSS selected sources in $g$, $r$, $i$, $u$, and $z$ bands for panels \textbf{a}--\textbf{e}, respectively.}
	\label{fig:SNumCounts}
\end{figure*}

Further confirmation of the basic accuracy of the model used to predict the EBL can be found from comparing observed source flux counts to predicted ones across different wavelengths. 
Figures~\ref{fig:WNumCounts} and \ref{fig:SNumCounts} contain comparisons of the observed source flux histograms (black lines) to the predicted contribution of galaxies based on the mean LF of the posterior chains produced from the High $z$ Prior, High $z$ Trim Prior, and \WD\ samples (red, blue, and grey dashed lines, respectively), and a simple power law model fit to a subset of the data in each plot (orange dash-dotted line) that is, nominally `stars' (the subset range is highlighted in orange in the plots under the main one in the panel). 
The fit parameters are not reported here because they are beyond the scope of this work.
The top rows contain direct comparisons with multiple models, and the bottom rows show the fractional residual counts with Poisson uncertainties after subtracting off the sum of the stars model and the High $z$ Prior model (model counts in the denominator).

Figure~\ref{fig:WNumCounts} contains comparisons to the AllWISE flux counts for all sources within the northern galactic cap ($b \ge 30\deg$) that have no artifact flags set and a signal to noise ratio (SNR) at least $4$ in the plotted band, giving a total of $650$, $47$, and $4.7$ million sources in panels \textbf{a}--\textbf{c}, respectively.
The plotted flux is the standard point spread function (psf) flux, so it will have inaccuracies at the bright end. 
Those inaccuracies are unobservable, though, because no effort was made to separate stars from galaxies and the star counts dominate there. 
No completeness corrections were applied to any of the data, either, so the faint end of the observations is expected to undershoot the predictions.

Figure~\ref{fig:SNumCounts} contains counts from the Sloan Digital Sky Survey (SDSS) data release 10 database table named \code{PhotoObj}. 
The information in the plots shown in Figure~\ref{fig:SNumCounts} is nearly identical to the ones from Figure~\ref{fig:WNumCounts}, with the one major change being that the extinction correction in the SDSS columns \code{extinction\_}[band letter] were applied.
The SDSS sources are limited to two circular regions with $9\deg$ radius that are nearly antipodes -- centered at J2000 right ascension and declination $(163.56309\deg, 7.27216\deg)$ and $(343.56309\deg, -1.27216\deg)$.
A source is excluded if it has any of the following flags, as explained in the SDSS database schema browser\footnote{\url{http://skyserver.sdss.org/dr10/en/help/browser/browser.aspx\#\&\&history=enum+PhotoFlags+E}}, set for the band in question (column named \code{flags\_}[band letter]): \code{EDGE}, \code{BLENDED}, \code{NODEBLEND}, \code{SATURATED}, \code{TOO\_LARGE}, \code{MOVED}, \code{MAYBE\_CR}, and \code{MAYBE\_EGHOST}.
Each source also had to have an $\mathrm{SNR}\ge 4$, just like the AllWISE sources, giving a total of $6.4$, $8.5$, $9.0$, $1.5$, and $4.9$ million sources in panels \textbf{a}--\textbf{e}, respectively. 
The plotted fluxes are the \code{modelFlux\_}[band letter] columns, so they won't describe stars accurately, but that shouldn't matter for the same reason the psf fluxes do not meaningfully affect the AllWISE plots.

What the plots in Figures~\ref{fig:WNumCounts} and \ref{fig:SNumCounts} show is that the model used here performs better than expected in predicting the number counts at wavelengths where the Gaussian approximation of \Lsed\ is no longer valid (particularly SDSS $g$ and $u$). 
In all wavelengths and for all of the predictions plotted, the flux counts predictions are reasonably close to the observed histograms. 
The comparison makes clear how, using flux counts alone or using LFs measured at different wavelengths, it would be easy to get an estimate of the $3.4\micron$ background to be closer to the $5\pm 1 \kJy \sr^{-1}$ of the High $z$ Trim Prior based estimates, plotted in blue, than the High $z$ Prior based estimate, plotted in red, depending on the how strict the definition of `galaxy' is and how incompleteness at the faint end is modeled. 
The comparisons also show that there is still untapped information that can be used to more tightly constrain the full spectro-luminosity functional in future works. 
The over-prediction at the faint end of the W1 plot is to be expected because of incompleteness from both the limit of photometric sensitivity\footnote{\url{http://wise2.ipac.caltech.edu/docs/release/allwise/expsup/sec2_4a.html}} and confusion\footnote{\url{http://wise2.ipac.caltech.edu/docs/release/allsky/expsup/sec6_2.html\#brt\_stars}}, though the level of over-prediction for the High-$z$ Prior model at the faint end means it should be viewed as suspect for the purposes of predicting faint galaxy flux counts.

\section{Conclusion}
We showed from analyzing the mean SED of galaxies using more than a thousand galaxies with an average of more than 5 photometric observations of each galaxy, and the $2.4\micron$ LF of galaxies using more than half a million galaxies with spectroscopic redshifts, that the contribution of galaxies to EBL at $3.4\micron$ is $I_\nu = 9.0\pm 0.5\kJy \sr^{-1}$ ($\nu I_\nu = 8.0 \pm 0.4 \nW \meter^{-2} \sr^{-1} e\text{-fold}$), with systematic uncertainties unlikely to be greater than $2\kJy \sr^{-1}$. 
This value is consistent with both direct measures of the background and constraints based on blazar spectra. 
Recent work on the production of gamma rays by blazar protons relaxes the strength of these constraints \citep{Essey:2010,Essey:2011,Aharonian:2013}, leaving room for contributions from extended galaxy halos \citep[discussed in][]{Cooray:2012} and from a large faint galaxy population implied by the steepening faint end slope of the mass function \citep[compiled in][]{Conselice:2016}. 
Settling the contribution from high $z$ faint galaxies will require deeper redshift surveys, to more firmly establish the steepness of the faint end slope, and work to establish how the faint end of the LF cuts off. 
Two examples of possible mechanisms that can provide a faint end cutoff to the LF are: an intrinsic lower bound to the halo mass function, and a deviation in the halo mass to light conversion factor (caused, for example, by a lower limit on the halo mass capable of accreting and cooling baryons from the inter-galactic medium into star forming clouds).

\section{Acknowledgements}
We would like to thank the anonymous reviewer for the helpful suggestions that improved the quality of this work.

This publication makes use of data products from the Wide-field Infrared Survey Explorer, which is a joint project of the University of California, Los Angeles, and the Jet Propulsion Laboratory/California Institute of Technology, and NEOWISE, which is a project of the Jet Propulsion Laboratory/California Institute of Technology. WISE and NEOWISE are funded by the National Aeronautics and Space Administration.

Funding for SDSS-III has been provided by the Alfred P. Sloan Foundation, the Participating Institutions, the National Science Foundation, and the U.S. Department of Energy Office of Science. The SDSS-III web site is http://www.sdss3.org/.

SDSS-III is managed by the Astrophysical Research Consortium for the Participating Institutions of the SDSS-III Collaboration including the University of Arizona, the Brazilian Participation Group, Brookhaven National Laboratory, Carnegie Mellon University, University of Florida, the French Participation Group, the German Participation Group, Harvard University, the Instituto de Astrofisica de Canarias, the Michigan State/Notre Dame/JINA Participation Group, Johns Hopkins University, Lawrence Berkeley National Laboratory, Max Planck Institute for Astrophysics, Max Planck Institute for Extraterrestrial Physics, New Mexico State University, New York University, Ohio State University, Pennsylvania State University, University of Portsmouth, Princeton University, the Spanish Participation Group, University of Tokyo, University of Utah, Vanderbilt University, University of Virginia, University of Washington, and Yale University.

This research has made use of the NASA/ IPAC Infrared Science Archive, which is operated by the Jet Propulsion Laboratory, California Institute of Technology, under contract with the National Aeronautics and Space Administration.

RJA was supported by FONDECYT grant number 1191124.

Funded by Chinese Academy of Sciences President’s International Fellowship Initiative. Grant No. 2019PM0017.

This project is partially supported by the CAS International Partnership Program No.114A11KYSB20160008 and CAS Interdisciplinary Innovation Team.

\bibliography{ThesisBib}

\end{document}